\documentclass[preprint,12pt]{elsarticle}
\journal{Nuclear Instruments and Methods} 
\usepackage{graphicx}
\usepackage{amssymb}

\newcommand{\etal}{{\it et al.}}
\newcommand{\phosgain}{90}
\newcommand{\ptopgain}{7} 

\graphicspath{
  {./110505/}
}

\begin{document}
\begin{frontmatter}

\title{Performance of a 20-in. photoelectric lens image intensifier tube}
\author{Yoichi~Asaoka and Makoto~Sasaki} 
\address{Institute for Cosmic Ray Research, University of Tokyo, 
	 5-1-5 Kashiwanoha, Kashiwa 277--8582 Japan}

\begin{abstract}
We have evaluated a 20-in. photoelectric lens image intensifier tube (PLI) to 
be mounted on the spherical focal surface of the Ashra light collectors, where
Ashra stands for All-sky Survey High Resolution 
Air-shower Detector, 
an unconventional optical collector complex
that images air showers produced by very high energy cosmic-ray particles 
in a 42$^\circ$-diameter field of view 
with a resolution of a few arcminutes.
The PLI, the worlds largest image intensifier, 
has a very large effective photocathode area of 20-in.
diameter and reduces an image size to less than 1-inch diameter using the
electric lens effect. 
This enables us to use a solid-state imager to take focal surface
images in the Ashra light collector. 
Thus, PLI is a key technology for the Ashra experiment to realize
a much lower pixel cost in comparison with other experiments using
photomultiplier arrays at the focal surface.
In this paper we present the design and performance 
of the 20-in. PLI.
\end{abstract}

\begin{keyword}
First generation image intensifier tube;
Large sensitive area;
Photodetector;
High energy astrophysics;
Ashra Experiment
\end{keyword}

\end{frontmatter}

\section{Introduction}
The All-sky Survey High Resolution Air-shower
detector (Ashra)~\cite{ASHRAoptics,ASHRAiit,Ashra}
is an experiment for obtaining fine images of air-showers 
produced by very high energy (VHE) cosmic-ray particles 
as well as directly observing starlight to monitor
optical transients.
Ashra has an potential to provide a systematic 
exploration of extragalactic VHE particle radiators in the
universe.

The Ashra observational station consists of 12 
light collectors covering 77\% of the entire sky
with a total of 50 mega pixels in CMOS sensor arrays.
Each light collector has a 42$^\circ$ field of view (FOV) and
a spot size resolution of a few arcmin.
The focal surface of the Ashra optical system consists of a
photoelectric lens image intensifier tube (PLI) sensitive to UV and
visible rays. 
The PLI reduces an image to the size of a solid state imaging device, such as
a CMOS sensor, using the electric lens effect.
Since this feature drastically reduces the pixel cost compared to
experiments using photomultiplier tube (PMT) arrays~\cite{HESS,MAGIC,AugerFluo}, it is a key
concept in the ability of Ashra to perform all-sky surveys with a few arcmin resolution. 
Details of the design and performance of PLI are described in this
paper.

\section{General Description}
\begin{figure}[bth!]
\centering
	\includegraphics[width=4in]{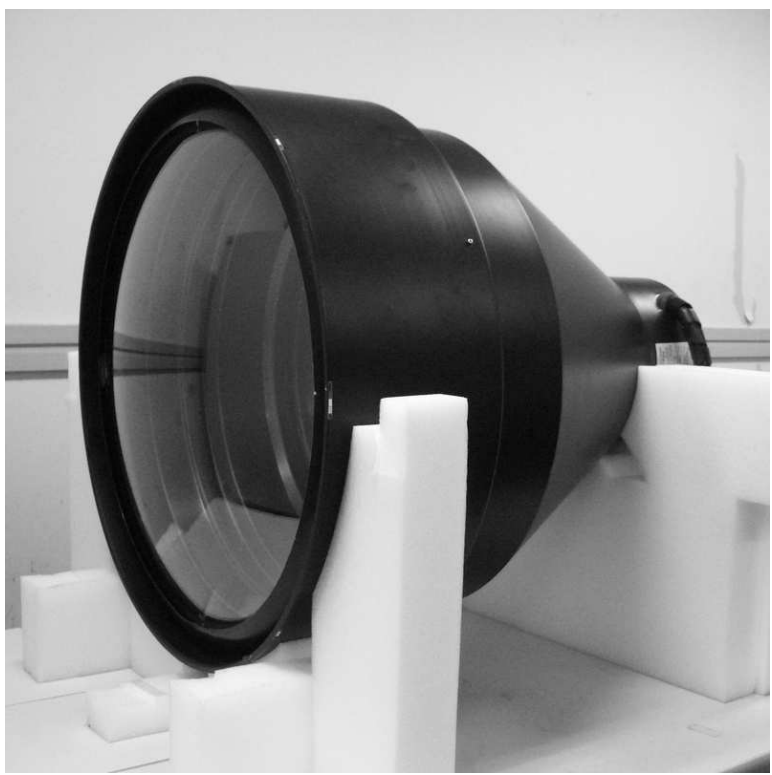} 
\caption{A photograph of a 20-in. photoelectric lens image intensifier tube (PLI).}
\label{fig:photo}
\end{figure}

The PLI is a 1st generation image intensifier tube, sensitive to
UV and visible photons,
fabricated by Toshiba Electron Tubes \& Devices Co., Ltd (TETD). 
Fig.~\ref{fig:photo} shows a photograph of a 20-in. PLI.

The PLI features are as follows:
\begin{enumerate}
\item the worlds largest image intensifier,
\item large input diameter of 500~mm,
\item small output diameter of $<$25~mm,
\item high imaging capability of 2~Lp/mm at input window,
\item very fast response using a P47 phosphor screen,
\item good single photoelectron detection capability, 
\item photon-to-photon amplification factor of $\sim$\ptopgain. 
\end{enumerate}

The 20-in. PLI is based on the success on the development of a 16-in. UV-ray image intensifier tube
\cite{UVII}, which is a modification of a commercial X-ray image intensifier
widely used in medical imaging. 16-in. image intensifier tubes were already used in 
our observation using prototype detectors \cite{GCN2846,Ashra-ICRC07}.
The PLI consists of three main components; a 20-in. input window, main parts of vacuum 
tube containing several electrodes, and an output window.
The input window is a UV transparent
glass, sensitive to UV and visible photons (not a thin metal plate, as is used in X-ray image intensifiers).
The input window has a spherical shape with the proper curvature radius to
work as the focal surface of the Ashra light collector.
Photoelectrons generated at a photocathode on the inner
surface of the input window are drawn toward the output window by an
electric field produced by electrodes. 
The voltage applied to each electrode was optimized to get 
a uniform resolution over the whole effective area of PLI.
The anode potential of the electric lens is $\sim$40~kV. 
The input image is reduced by a factor of 20 or more as a result of the electric lens effect.
The concave-shaped output focal surface of the electric lens is formed of a  
phosphor material screen on a concave-shaped fiber optic
plate (FOP).
As a phosphor material we used Y$_2$SiO$_5$:Ce, known as P47, to
take advantage of its very fast 10\% decay constant of $\sim$110~ns.
An incident photoelectron is amplified to $\sim$\phosgain~photons at the
phosphor screen.
Finally, a planar image is obtained on the output side of the FOP window.

\section{Performance}
Ashra-1 collaboration already started the observation using some of the 
finished 20-in. PLIs as reported in Refs.~\cite{ML-ICRC07,GCN8632,CNeu-ICRC09,OPF-ICRC09,GCN11291}.
Qualitatively, similar results to the performance described in this paper
were obtained with PLIs used in the observation.
Here, we quantitatively evaluated various performance parameters
using the latest PLI:
the quantum efficiency, phosphor gain and imaging capability.
In the following, each item is described in detail.

\subsection{Quantum efficiency}
\begin{figure}[hbt!]
\centering
\includegraphics[width=5in]{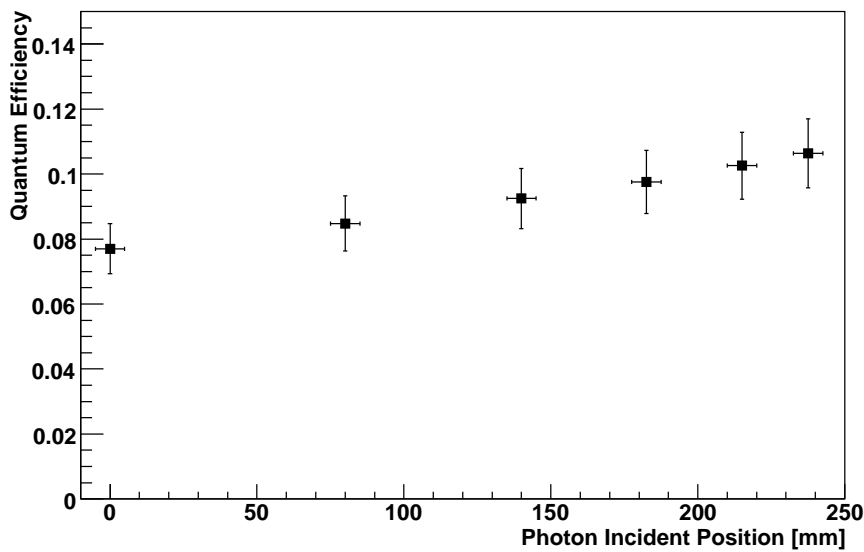}
\caption{Quantum efficiency of the 20-in. PLI as a function of
photon incident position. 0 and 250~mm correspond to the
center and the edge of the effective input area, respectively.}
\label{fig:qe}
\end{figure}

At first, we evaluated the sensitivity of the PLI.
Fig.~\ref{fig:qe} shows the quantum efficiency as a
function of photon incident position (where the error bars 
represent the quadratic sum of the statistical and systematic errors). 
The quantum efficiency was calculated as a sensitivity relative to a calibrated PMT.
A UV LED with a wavelength peaked at 365~nm was used as a light source.
First, we measured the photocathode current of the calibrated PMT where the voltage of 250~V was 
applied between photocathode and dynodes. 
All the dynodes and the anode were
connected and resultant current saturated at much lower voltage.
Second, we measured the photocathode current of the PLI where all the electrodes and the anode
were connected and the voltage of 250~V was applied between photocathode and electrodes.
We confirmed that the measured current was almost saturated at 250~V.
Since the obtained current ratio corresponds to the ratio of the quantum efficiencies,
the quantum efficiency of the PLI was calculated for each input position.
The dominant uncertainty in the quantum efficiency was
the uncertainty in the measurements of the quantum
efficiency for the calibrated PMT. 

As shown in Fig.~\ref{fig:qe}, 
the quantum efficiency was measured to be 7.7\% at the center
of the input window and was increasing toward the edge. 
At the edge, the quantum efficiency was measured to be 10.5\%.
The sensitivity was confirmed using pulsed YAG laser (peak wavelength of 355nm);
we obtained the quantum efficiency consistent with the results described above at
the center region.

\subsection{Phosphor gain}
\begin{figure}[hbt!]
\centering
\includegraphics[width=5in]{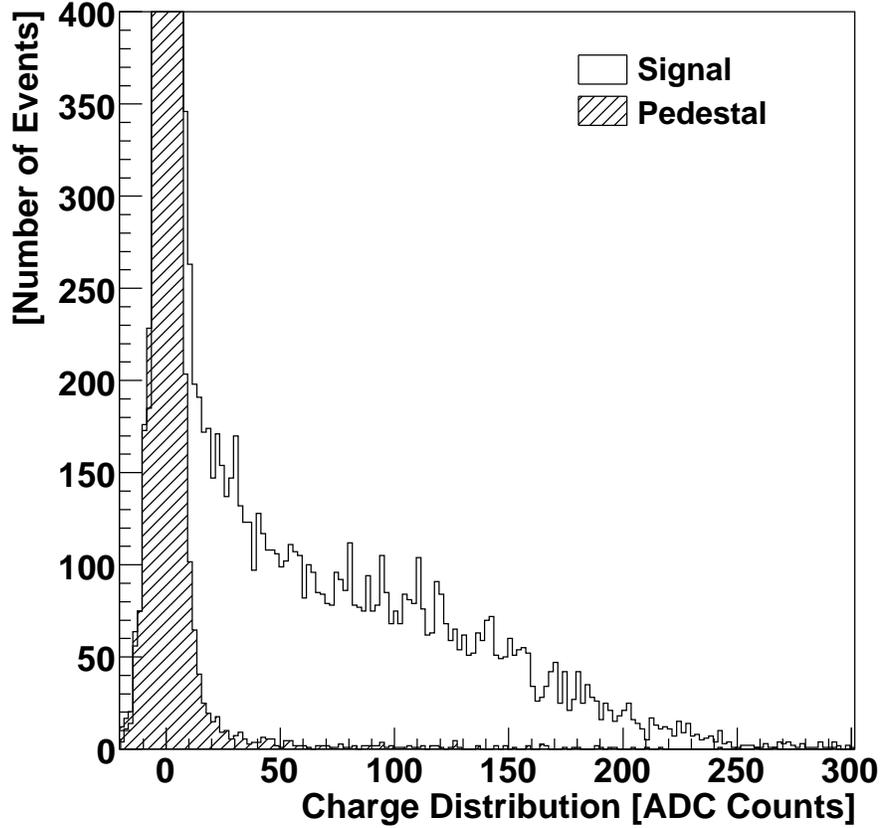}
\caption{Single photoelectron distribution.
The open and hatched histograms represent the signal and pedestal
distributions, respectively.}
\label{fig:spedst}
\end{figure}

In order to evaluate the performance of the phosphor screen, we investigated
the single photoelectron response.
To do this, we used a LED illuminated by pulsed input voltage.
The output photons from LED was set to 0.1 photoelectrons per pulse on average, 
for which we could neglect a signal of two photoelectrons or more.
As a result, we obtained a pure one photoelectron signal.
The gain properties of the output screen were measured with a PMT
directly coupled to the output FOP of the PLI. 
Fig.~\ref{fig:spedst} shows the photon distribution emitted from the
output screen for single photoelectron incidence.
The hatched area corresponds to the pedestal distribution.
We found that about \phosgain~photons were emitted on average. 
An acceptable structure was obtained for the single
photoelectron distribution.

\begin{figure}[hbt!]
\centering
\includegraphics[width=5in]{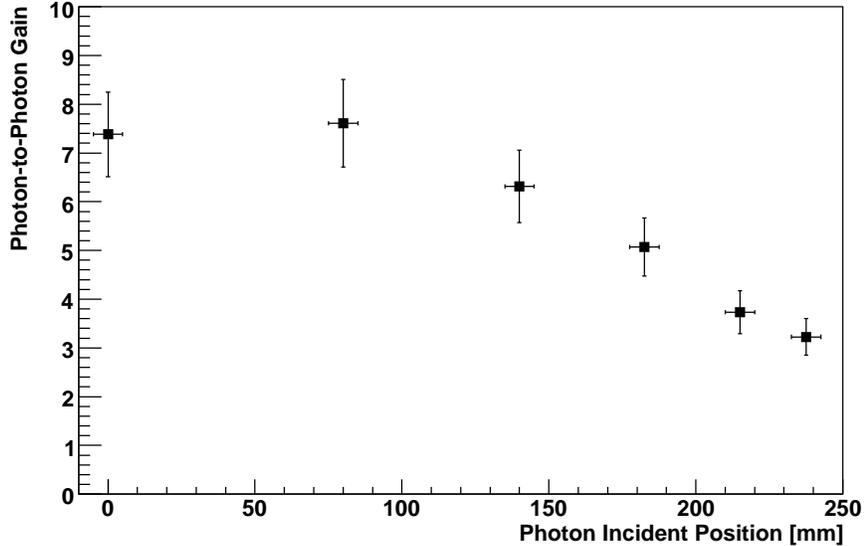}
\caption{Photon-to-photon gain of the 20-in. PLI as a function of
photon incident position. 0 and 250~mm correspond to the
center and the edge of the effective input area, respectively.}
\label{fig:p2p}
\end{figure}

To evaluate photon-to-photon gain of PLI, on the other hand, we used a very stable YAP pulser
(YAlO$_{3}$:Ce+241Am, peak wavelength of 370~nm, and $e^{-1}$ decay time of 28~ns).
At first, photon intensity of the YAP pulser was measured by the calibrated PMT.
Then, YAP pulser was placed at the several positions of the input window of the PLI.
The output photons were measured with the same PMT
directly coupled to the output window of the PLI. 
The resultant photon-to-photon gains were obtained as the ratio of the 
PMT signals after correcting sensitivity difference between YAP and P47 spectrum,
and are plotted in Fig.~\ref{fig:p2p} 
as a function of the photon incident position. 
The error bars represent the quadratic sum of the statistical and
systematic errors.
The most significant contribution to the overall error was the
uncertainty in the P47 spectrum, which 
might cause the sensitivity difference of the calibrated PMT.

The obtained photon-to-photon gain includes the following effects:
\begin{enumerate}
\item input glass transmittance,
\item quantum efficiency,
\item collection efficiency of photoelectrons,
\item efficiency of the output phosphor screen,
\item averaged photon output per single photoelectron incidence.
\end{enumerate}
In the previous section, the measured quantum efficiency only accounts for the 
input glass transmittance and the raw quantum efficiency (points 1 and 2 above),
which is the same definition as the general use.
In photon-to-photon gain measurements, however, 
the efficiency of the photoelectron collection and of the output screen
are also included. 
Combined with the single photoelectron distribution measurement shown in Fig.~\ref{fig:spedst}, 
we confirmed that these efficiencies (points 3 and 4 above) are reasonably high. 

Note, however, that there are several deficits in the peripheral region of the output phosphor screen 
of the PLI tested in this paper.
The brightness of the phosphor screen in those deficits was much worse
(1/2 or less) than that of normal area.
The radial scan of the photon-to-photon gain was performed along the direction 
where no such deficits in the output phosphor screen was seen in order to 
demonstrate the normal performance expected in the PLI design.

\begin{figure}[hbt!]
\centering
\includegraphics[width=5in]{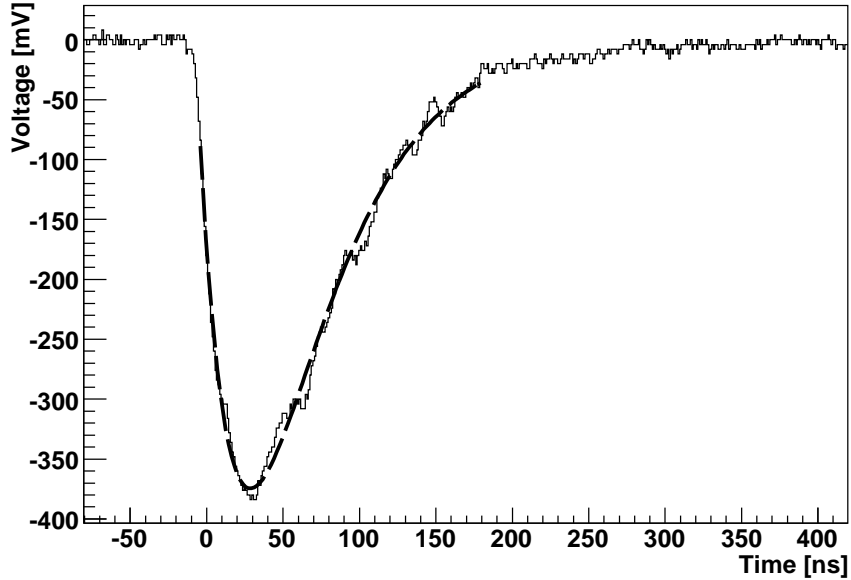}
\caption{A typical example of YAP signal taken through the 20-in. PLI.
	The 10\% decay time of the P47 screen was determined to be
	110~ns from the fit to the data shown as long dashed line (see text).}
\label{fig:decay}
\end{figure}

Fig.~\ref{fig:decay} shows an typical waveform of the YAP signal taken through the 
PLI. It represents the convolution of two decay times of the YAP and the
P47 phosphor screen. From the fit to the data shown as the long dashed 
line in Fig.~\ref{fig:decay} where the YAP decay time was fixed to 28~ns ($e^{-1}$),
the 10\% decay time of the P47 screen was determined to be
110~ns, which is consistent with the typical catalog value. 
This very fast decay feature enables us to detect cosmic-ray air-light
signals with a high signal-to-noise ratio.

\subsection{Imaging capability}
The point spread function (PSF) was next measured to evaluate
the PLI resolution quantitatively.
A UV-blue laser was incident on the PLI through a 10-$\mu$m pinhole 
and a focusing lens,
and the output image was taken by an 6-mega-pixel digital camera.
The laser was mounted on a 2-axis positioning stage such that its direction was 
parallel to the PLI axis, where the PLI axis is defined as a line between
the center of the input window and the center of the output window.
By varying the laser incident position 
in the plane perpendicular to the PLI axis, 
we measured the PLI resolution as a function of photon incident
position.
The spot size of the laser was measured before evaluating the PLI resolution.
We found that the full width half maximum (FWHM) of the laser spot was within 
120~$\mu$m in the region where the PLI photocathode would be placed. Note that 
this region was about 45~mm in length because of the spherical shape of the PLI 
input window.
At the input window of PLI, 1~arcmin resolution corresponds to 200~$\mu$m.
Considering the fact that FWHM is 2.35~$\sigma$ in the case of a Gaussian distribution,
FWHM of 470~$\mu$m corresponds to 1~arcmin resolution
by using a 1-$\sigma$ deviation as the measure of the resolution.  
Thus, we concluded that our laser system could be used as the light source of 
PSF measurements since obtained FWHMs of $<$120~$\mu$m were well 
below 1~arcmin resolution at the input window.

\begin{figure*}[hbtp!]
\centering
 \begin{tabular}{ccc}
  \begin{minipage}{0.45\hsize}
   \begin{center}
	\includegraphics[width=2.2in]{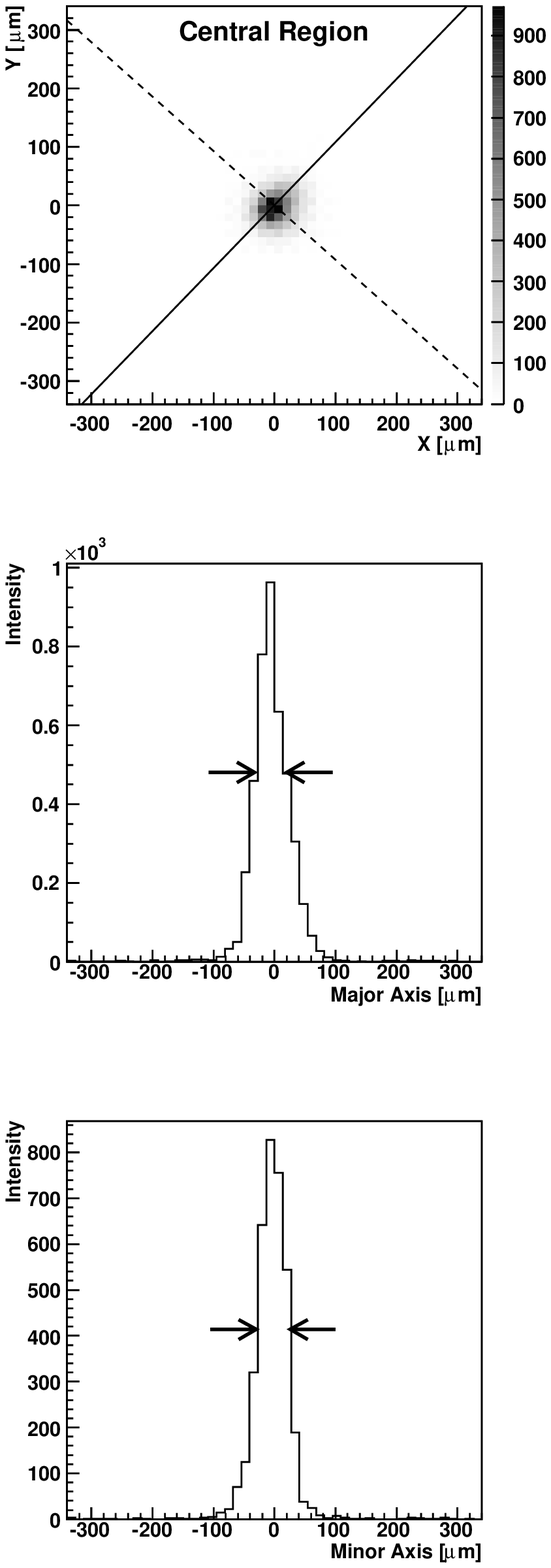}
   \end{center}
  \end{minipage} &
  \begin{minipage}{0.06\hsize}
	\vspace*{0.2cm}
  \end{minipage} &
  \begin{minipage}{0.45\hsize}
   \begin{center}
	\includegraphics[width=2.2in]{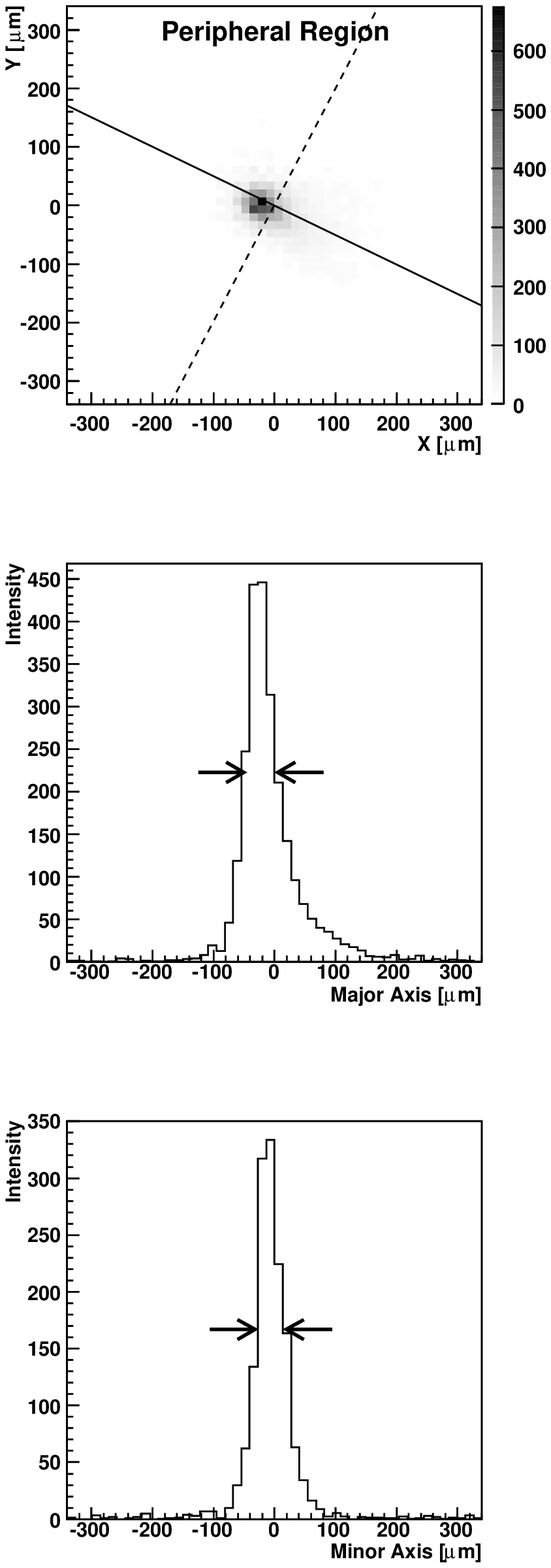}
   \end{center}
  \end{minipage}\\
 \end{tabular}
\caption{Examples of point spreads of the 20-in. PLI: ({\it left})
central region, ({\it right}) peripheral region.
From top to bottom the panels show a spot diagram, the cross
section along the major axis, and the cross
section along the minor axis.
In the spot diagrams, solid and dashed lines represent major and minor axes, 
respectively.
FWHMs are indicated by arrows in the cross-sections.
}
\label{fig:psf}
\end{figure*}

Fig. \ref{fig:psf} shows examples of PSF spot images taken with the above explained 
laser system;
the left and right panels show spot images in the central and peripheral
regions, respectively. 
From top to bottom the panels show a spot diagram, the cross
section along the major axis (determined from the elliptical shape of the spot), 
and the cross-section along the minor axis (taken with the axis perpendicular 
to the major axis). 
A good resolution of FWHM$\sim$50~$\mu$m was achieved at the central
region, 
though there was some deterioration at the 
peripheral regions due to an increase in the tails of the
distributions.
In those spot diagrams, radial directions correspond to $-$45$^\circ$.
It should be noted that the directions of the tail in the peripheral regions
tend to turn away from the radial direction. Since the PLI should have 
revolution symmetry along its axis, this trend might suggest the asymmetry
in alignment of the parts and/or the other effects. 
Note also that the obtained resolution includes a small contribution from
the employed digital camera (one pixel corresponds to 13.6~$\mu$m considering the
magnification factor of close-up lens mounted on the camera).
A half of the pixel size was quadratically added to the quoted error 
to conservatively indicate the effect of pixelization
in the determination of FWHM.

\begin{figure}[hbt!]
\centering
\includegraphics[width=5in]{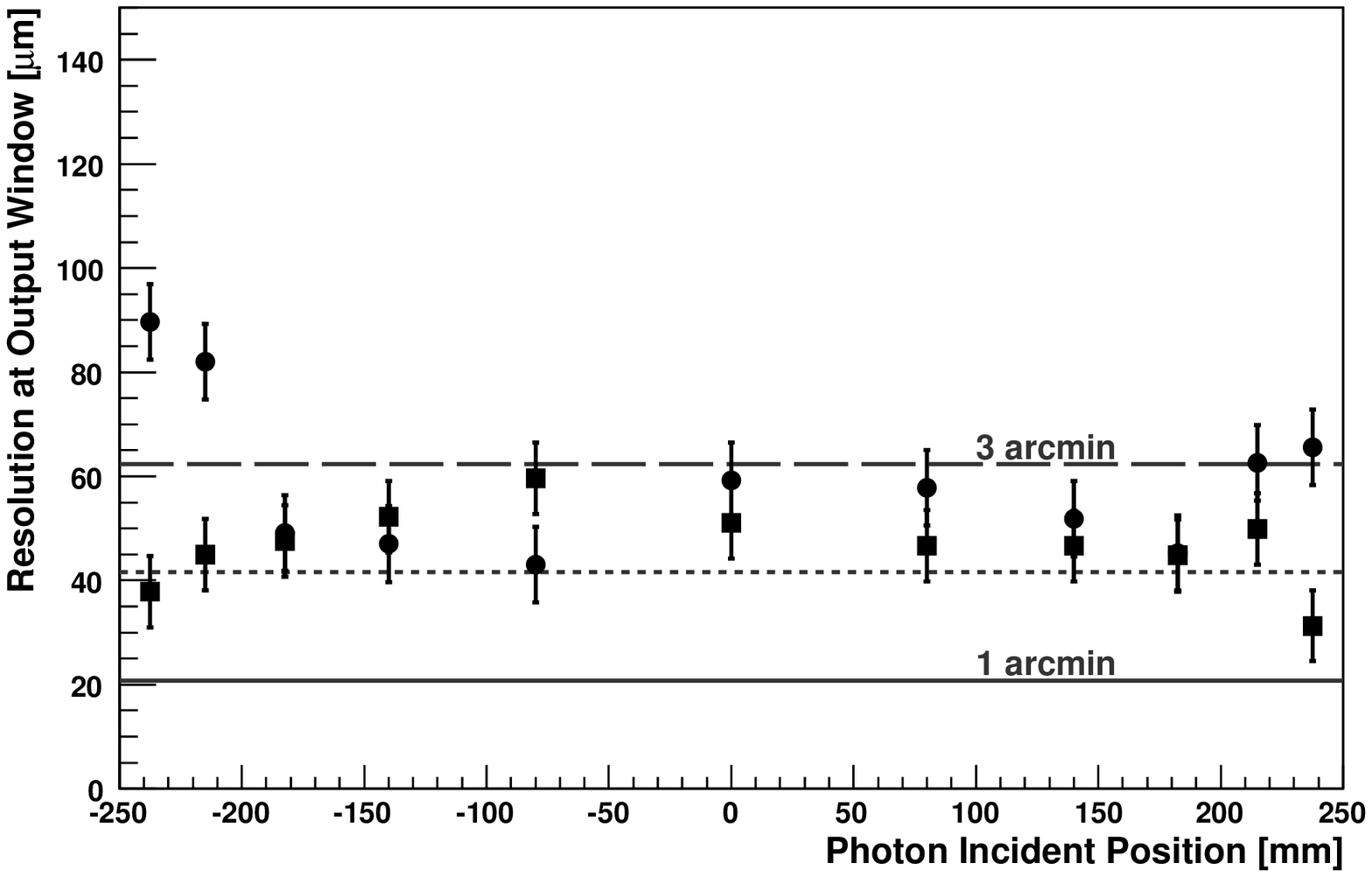}
\vspace*{0.15cm}
\caption{Resolution of the 20-in. PLI as a function of photon
incident position.
The closed circles and squares represent the FWHMs along the major and minor axes,
respectively.}
\label{fig:psf_dist}
\end{figure}

Fig.~\ref{fig:psf_dist} shows the PSF dependence on the photon
incident position.  
The closed circles and squares represent the FWHMs along the major and minor axes,
respectively. 
With the adopted electric field configuration, 
the magnification factor was measured to be 0.046 (0.045) at the center (edge),
resulting in the output image diameter of $\phi$~23mm.
We observed a good resolution corresponds to 2--3~arcmin in most region of the sensitive area.
However, there is an asymmetry between the left and right side 
resolutions, due possibly to the difference in the axis of the input, 
output window and the main parts of the PLI.

\section{Conclusion}
We have succeeded in producing a 20-in. diameter photoelectric lens image 
intensifier tube (PLI) for the first time.
The PLI is designed to be mounted on the focal surface of the Ashra
optical system.
The PLI reduces the image size into solid state imaging devices with some 
amplification.
It enables drastic pixel cost reduction compared to 
conventional photomultiplier arrays at the focal
surface of telescopes. 
Hence, the PLI is a key technology in the ability of Ashra to perform all-sky surveys
with a few arcmin resolution.
Furthermore, because of its large diameter and fine imaging capability,
PLIs may have many other applications in high-energy and
astrophysical experiments. 

\section*{Acknowledgment}
This work was mostly supported by the Coordination of Science and Technology 
(157-20004100), and was supported in part by the Grant-in-Aid 
for Scientific Research (16740130) of the Ministry of
Education, Science, Sports and Culture in Japan.

\end{document}